\documentclass{emulateapj}

\newcommand{\kms}{\ifmmode{\rm km\thinspace s^{-1}}\else km\thinspace s$^{-1}$\fi}
\newcommand\aq{AQ\,Ser}

\shorttitle{\aq}
\shortauthors{Torres et al.}
\begin{document}

\title{Absolute properties of the eclipsing binary system AQ
Serpentis: A stringent test of convective core overshooting in stellar
evolution models}

\author{
Guillermo Torres\altaffilmark{1},
Luiz Paulo R.\ Vaz\altaffilmark{2},
Claud H.\ Sandberg Lacy\altaffilmark{3}, and
Antonio Claret\altaffilmark{4}
}

\altaffiltext{1}{Harvard-Smithsonian Center for Astrophysics, 60
Garden Street, Cambridge, MA 02138, USA, e-mail:
gtorres@cfa.harvard.edu}

\altaffiltext{2}{Depto.\ de F\'{\i}sica, ICEx-UFMG, C.P. 702,
  30.161-970 Belo Horizonte, MG, Brazil, e-mail: lpv@fisica.ufmg.br}

\altaffiltext{3}{Department of Physics, University of Arkansas,
Fayetteville, AR 72701, USA, e-mail: clacy@uark.edu}

\altaffiltext{4}{Instituto de Astrof\'\i sica de Andaluc\'\i a, CSIC,
Apartado 3004, 18080 Granada, Spain; e-mail: claret@iaa.es}

\begin{abstract}

We report differential photometric observations and radial-velocity
measurements of the detached, 1.69-day period, double-lined eclipsing
binary \aq. Accurate masses and radii for the components are
determined to better than 1.8\% and 1.1\%, respectively, and are $M_1
= 1.417 \pm 0.021\,M_{\sun}$, $M_2 = 1.346 \pm 0.024\,M_{\sun}$, $R_1
= 2.451 \pm 0.027\,R_{\sun}$, and $R_2 = 2.281 \pm
0.014\,R_{\sun}$. The temperatures are $6340 \pm 100$\,K (spectral
type F6) and $6430 \pm 100$\,K (F5), respectively. Both stars are
considerably evolved, such that predictions from stellar evolution
theory are particularly sensitive to the degree of extra mixing above
the convective core (overshoot). The component masses are different
enough to exclude a location in the H-R diagram past the point of
central hydrogen exhaustion, which implies the need for extra mixing.
Moreover, we find that current main-sequence models are unable to
match the observed properties at a single age even when allowing the
unknown metallicity, mixing length parameter, and convective
overshooting parameter to vary freely and independently for the two
components. The age of the more massive star appears systematically
younger. \aq\ and other similarly evolved eclipsing binaries showing
the same discrepancy highlight an outstanding and largely overlooked
problem with the description of overshooting in current stellar
theory.

\end{abstract}


\section{Introduction}
\label{sec:introduction}

The eclipsing binary \aq\ (GSC~00340-00588, BD+03~3015, $V = 10.58$)
was discovered as a variable star by \cite{Hoffmeister1935}. Its
correct period of 1.6874 days and eclipse ephemeris were determined
much later by \cite{Soloviev1951}. The light curve shows relatively
deep (0.6 mag) and nearly identical primary and secondary eclipses,
and the spectral types of the stars have been reported as F5 and A2
\citep{Hill:75}, although the latter classification for the less
massive star is probably too early.  The system has been little
studied since its discovery, other than the occasional measurement of
times of eclipse.

The main motivation for this work is to present new photometric and
spectroscopic observations of \aq, with which we determine for the
first time accurate absolute dimensions for the system and establish
the evolutionary status of the stars. Both components appear to be at
the very end of their hydrogen-burning phase, a location of the H-R
diagram in which only a few other well-measured eclipsing binaries are
found. The predicted properties of such stars from stellar evolution
theory are especially sensitive to the degree of convective core
overshooting adopted in the models, and a previous study by
\cite{Clausen:10} has highlighted the difficulties that current models
appear to have in reproducing the stellar properties of both
components at a single age. The newly determined absolute dimensions
for \aq\ allow us an opportunity to revisit this problem here.

\section{Observations and Reductions}
\label{sec:observations}

\subsection{Differential photometry}
\label{sec:photometry}

Photometric measurements of \aq\ were determined by two different and
independent robotic observatories: the URSA WebScope, and the NFO
WebScope.  The URSA WebScope uses a 10-inch Meade LX200 SCT telescope
with an SBIG ST8 CCD camera, housed in a Technical Innovations
RoboDome on the roof of the Kimpel Hall on the University of Arkansas
campus at Fayetteville, and is controlled by an Apple Macintosh G4
computer in a nearby control room.  The field of view is about
20$\times$30 arc minutes.  Observations with a Bessel $V$ filter were
carried out from 2003 June to 2011 July, producing a total of 8642
science frames from 80-second exposures.  The two comparison stars for
\aq\ (`var'), both within 8 arc minutes of the variable star, were
GSC~00340-00252 (`comp'; $V = 10.99$, \ion{G5}{5}) and GSC~00341-00211
(`ck'; $V = 11.60$, \ion{G2}{5}).  It was eventually found that the ck
star is a low-amplitude variable with a sinusoidal variation of
half-amplitude 0.017 mag and a period of about four years.
Differential magnitudes in this study were therefore based on the
var$-$comp magnitudes only.

The NFO WebScope is located near Silver City (NM) in a roll-off roof
structure, and consists of a 24-inch Cassegrain reflector with a
field-widening correcting lens near the focus
\citep[see][]{GrauerNeelyLacy:08}.  At the focus is a camera based on
the Kodak KAF-4301E CCD chip, with a field of view of about
27$\times$27 arc minutes.  \aq\ was observed at the NFO between 2005
January and 2007 June, producing a total of 6694 observations from
80-second exposures with a Bessel $V$ filter.

All images were measured using a computer application ({\tt Measure})
that matched a pattern file with the image, and then determined the
differential magnitude after correction for dark current, sky
brightness, and responsivity variations across the field of view.

As we have noted in the past \citep[e.g.,][]{Lacy:08}, the telescopes
we used in this study produce systematic shifts of a few hundredths of
a magnitude in the photometric zero point from night to night, and in
the case of the NFO WebScope, from one side of the German equatorial
mount axis to the other.  The shifts are very much less for the URSA
WebScope than for the NFO, which shows that this is an effect of the
optical system being used, and is not intrinsic to the stars
themselves.  The offsets are due to a non-uniform responsivity across
the field of view, combined with imprecise centering from night to
night.  In the case of the NFO, we removed most of this effect by
using dithered exposures of open clusters to fit a 2-D polynomial
function to the responsivity variations, resulting in a photometric
flat that is included in the initial data reduction procedures.
Residual offsets remaining after this process were then removed by
using an initial photometric orbital fit model (see
Sect.~\ref{sec:LCanalysis}) to determine the values of the nightly
offsets and to remove them from the data.  In this case, 130 nightly
shifts were removed from the URSA data, and 197 shifts were removed
from the NFO data.  The typical precision of the final \aq\ data sets
is about 9 mmag for URSA and 5 mmag for NFO.  The measurements
including nightly corrections are listed in Table~\ref{tab:ursa}
(URSA) and Table~\ref{tab:nfo} (NFO).

\begin{deluxetable}{ccc}
\tablecaption{Differential $V$-band measurements of \aq\
from the URSA WebScope.\label{tab:ursa}}
\tablewidth{220pt}
\tablehead{
\colhead{HJD} & 
\colhead{Orbital} & 
\colhead{$\Delta V$} \\
\colhead{($2,\!400,\!000+$)} &
\colhead{phase} &
\colhead{(mag)}
}
\startdata
 52814.60641 &  0.0963 &  $-$0.366 \\
 52814.60833 &  0.0975 &  $-$0.396 \\
 52814.61022 &  0.0986 &  $-$0.365 \\
 52814.61212 &  0.0997 &  $-$0.366 \\
 52814.61399 &  0.1008 &  $-$0.375 
\enddata
\tablecomments{Orbital phase is computed with the ephemeris in
Sect.~\ref{sec:ephemeris}. Table~\ref{tab:ursa} is published in its
entirety in the electronic edition of the journal. A portion is shown
here for guidance regarding its form and content.}
\end{deluxetable}

\begin{deluxetable}{ccc}
\tablecaption{Differential $V$-band measurements of \aq\
from the NFO WebScope.\label{tab:nfo}}
\tablewidth{220pt}
\tablehead{
\colhead{HJD} & 
\colhead{Orbital} & 
\colhead{$\Delta V$} \\
\colhead{($2,\!400,\!000+$)} &
\colhead{phase} &
\colhead{(mag)}
}
\startdata
 53377.03321 &  0.4000 &  $-$0.390 \\
 53377.03484 &  0.4010 &  $-$0.390 \\
 53377.03647 &  0.4019 &  $-$0.390 \\
 53377.03806 &  0.4029 &  $-$0.390 \\
 53377.03970 &  0.4038 &  $-$0.386 
\enddata
\tablecomments{Orbital phase is computed with the ephemeris in
Sect.~\ref{sec:ephemeris}. Table~\ref{tab:nfo} is published in its
entirety in the electronic edition of the journal. A portion is shown
here for guidance regarding its form and content.}
\end{deluxetable}

\subsection{Spectroscopy}
\label{sec:spectroscopy}

Spectroscopic observations of \aq\ were carried out at the
Harvard-Smithsonian Center for Astrophysics using an echelle
spectrograph on the 1.5-m Tillinghast reflector at the F.L.\ Whipple
Observatory (Mount Hopkins, AZ).  A single echelle order 45\,\AA\ wide
was recorded with an intensified photon-counting Reticon detector, at
a central wavelength near 5190\,\AA\ that includes the \ion{Mg}{1}\,b
triplet.  The resolving power of these observations is
$\lambda/\Delta\lambda \approx 35,\!000$.  We gathered 39 spectra
between 2004 March and 2008 June, with signal-to-noise ratios ranging
between 22 and 41 per resolution element of 8.5~\kms.

All our spectra appear double-lined. Radial velocities were obtained
using the two-dimensional cross-correlation technique TODCOR
\citep{Zucker:94}, with templates chosen from a large library of
calculated spectra based on model atmospheres by R.\ L.\ Kurucz
\citep[see][]{Nordstrom:94, Latham:02}. The four main parameters of
the templates are the effective temperature $T_{\rm eff}$, rotational
velocity ($v \sin i$ when seen in projection), metallicity [m/H], and
surface gravity $\log g$. The ones affecting the velocities the most
are $T_{\rm eff}$ and $v \sin i$. Consequently, we held $\log g$ fixed
at values of 4.0 for both stars, which is near the final values
reported below in Sect.~\ref{sec:absdim}, and we assumed solar
metallicity.  The optimum $T_{\rm eff}$ and $v \sin i$ values were
determined by running grids of cross-correlations, seeking the maximum
of the correlation coefficient averaged over all exposures and
weighted by the strength of each spectrum \citep[see][]{Torres:02}.
The rotational velocities we obtained are $v \sin i = 59 \pm 10\,\kms$
for the hotter and less massive star (hereafter star A) and $v \sin i
= 73 \pm 10\,\kms$ for the cooler one (star B).  The significant
rotational line broadening in both stars and the relatively low
signal-to-noise ratios cause the uncertainties above to be fairly
large, and also prevent us from establishing the temperatures
accurately.  Only a rough estimate of $T_{\rm eff}$ could be obtained.
The values adopted from our analysis in Sect.~\ref{sec:absdim} are
$T_{\rm eff} = 6430$\,K for the less massive component and $T_{\rm
eff} = 6340$\,K for the other. The uncertainty in these values has
little effect on the velocities.

We also determined the light ratio at the mean wavelength of our
observations (which is close to the $V$ band), following the
prescription by \cite{Zucker:94}. We obtained $\ell_{\rm B}/\ell_{\rm
A} = 1.05 \pm 0.04$, formally indicating that the cooler and more
massive star of the system is visually the brightest.

As in previous studies using similar spectroscopic material, we made
an assessment of potential systematic errors in our radial velocities
that may result from residual line blending as well as lines shifting
in and out of our narrow spectral window as a function of orbital
phase \citep[see][]{Latham:96}. We did this by performing numerical
simulations analogous to those described by \cite{Torres:97}, and we
applied corrections to the raw velocities based on these simulations
to mitigate the effect. The corrections were typically less than
2.5~\kms\ for the hotter star and less than 2~\kms\ for the cooler
star, which are smaller than our internal velocity errors
($\sim$5~\kms).  The effect of these corrections on the absolute
masses is minimal.

Finally, the stability of the zero-point of our velocity system was
monitored by taking nightly exposures of the dusk and dawn sky, and
small run-to-run corrections (typically under 1~\kms) were applied to
the velocities as described by \cite{Latham:92}. The adopted
heliocentric velocities including all corrections are listed in
Table~\ref{fig:RVs}, together with their uncertainties and the
residuals from our adopted orbital solution described below.

\begin{deluxetable*}{lcrrrr}[t!]
\tablecaption{Heliocentric radial velocities for \aq.\label{fig:RVs}}
\tablehead{
\colhead{HJD} & 
\colhead{Orbital} & 
\colhead{RV$_{\rm A}$} & 
\colhead{RV$_{\rm B}$} &
\colhead{$(O\!-\!C)_{\rm A}$} & 
\colhead{$(O\!-\!C)_{\rm B}$} 
\\
\colhead{($2,\!400,\!000+$)} &
\colhead{phase\tablenotemark{a}} &
\colhead{(\kms)} &
\colhead{(\kms)} &
\colhead{(\kms)} &
\colhead{(\kms)}
}
\startdata
  53073.9118  &   0.7651 & $  158.1 \pm 4.6 $  & $ -106.9 \pm 5.9 $ &   $ 10.8  $ & $  -7.2 $ \\ 
  53096.9083  &   0.3932 & $  -62.0 \pm 4.2 $  & $   90.6 \pm 5.3 $ &   $ -3.4  $ & $  -5.0 $ \\ 
  53103.9912  &   0.5907 & $   87.1 \pm 4.8 $  & $  -36.6 \pm 6.2 $ &   $ -2.2  $ & $   8.0 $ \\ 
  53126.9531  &   0.1983 & $ -103.8 \pm 3.8 $  & $  128.5 \pm 4.8 $ &   $ -3.8  $ & $  -6.6 $ \\ 
  53133.8183  &   0.2667 & $ -110.8 \pm 4.2 $  & $  153.4 \pm 5.4 $ &   $ -4.8  $ & $  12.7 $ \\  
  53154.8085  &   0.7059 & $  142.7 \pm 3.5 $  & $ -101.2 \pm 4.4 $ &   $ -0.2  $ & $  -5.6 $ \\  
  53155.7765  &   0.2795 & $ -105.8 \pm 3.3 $  & $  135.4 \pm 4.2 $ &   $ -1.2  $ & $  -4.0 $ \\  
  53160.7788  &   0.2440 & $ -104.6 \pm 4.8 $  & $  147.3 \pm 6.0 $ &   $  2.0  $ & $   6.0 $ \\  
  53161.7409  &   0.8141 & $  141.8 \pm 4.6 $  & $  -97.0 \pm 5.9 $ &   $  4.1  $ & $  -6.4 $ \\  
  53182.7419  &   0.2597 & $ -109.3 \pm 5.2 $  & $  141.0 \pm 6.6 $ &   $ -2.9  $ & $  -0.2 $ \\  
  53188.6973  &   0.7889 & $  140.4 \pm 3.1 $  & $  -90.3 \pm 3.9 $ &   $ -3.7  $ & $   6.4 $ \\  
  53192.7251  &   0.1759 & $  -88.6 \pm 5.0 $  & $  137.3 \pm 6.4 $ &   $  4.5  $ & $   8.8 $ \\  
  53452.8657  &   0.3396 & $  -84.2 \pm 4.8 $  & $  126.3 \pm 6.0 $ &   $  2.9  $ & $   3.6 $ \\  
  53455.9574  &   0.1718 & $  -96.9 \pm 5.2 $  & $  122.6 \pm 6.6 $ &   $ -5.2  $ & $  -4.5 $ \\  
  53456.9539  &   0.7624 & $  152.4 \pm 5.1 $  & $ -101.0 \pm 6.5 $ &   $  4.9  $ & $  -1.2 $ \\  
  53483.8617  &   0.7084 & $  136.2 \pm 4.6 $  & $ -100.5 \pm 5.9 $ &   $ -7.3  $ & $  -4.4 $ \\  
  53488.8435  &   0.6607 & $  127.1 \pm 3.8 $  & $  -72.0 \pm 4.8 $ &   $ -1.2  $ & $   9.7 $ \\  
  53543.7496  &   0.1990 & $ -105.0 \pm 4.8 $  & $  133.5 \pm 6.1 $ &   $ -4.8  $ & $  -1.7 $ \\  
  53576.7161  &   0.7355 & $  152.8 \pm 5.0 $  & $  -97.1 \pm 6.4 $ &   $  5.5  $ & $   2.6 $ \\  
  53866.8395  &   0.6675 & $  129.1 \pm 5.1 $  & $  -85.0 \pm 6.4 $ &   $ -2.1  $ & $  -0.6 $ \\  
  53872.8047  &   0.2026 & $  -87.0 \pm 5.0 $  & $  125.4 \pm 6.4 $ &   $ 14.1  $ & $ -10.7 $ \\  
  53873.7141  &   0.7415 & $  145.0 \pm 4.6 $  & $  -99.6 \pm 5.9 $ &   $ -2.6  $ & $   0.5 $ \\  
  53895.7489  &   0.7997 & $  138.6 \pm 4.6 $  & $  -91.5 \pm 5.8 $ &   $ -3.0  $ & $   2.9 $ \\  
  53901.6867  &   0.3186 & $  -94.6 \pm 5.3 $  & $  127.6 \pm 6.7 $ &   $  0.4  $ & $  -2.7 $ \\  
  54137.0002  &   0.7693 & $  150.5 \pm 5.1 $  & $  -97.7 \pm 6.4 $ &   $  3.6  $ & $   1.7 $ \\  
  54158.9724  &   0.7904 & $  139.8 \pm 5.7 $  & $  -94.1 \pm 7.2 $ &   $ -4.0  $ & $   2.3 $ \\  
  54162.9989  &   0.1766 & $  -93.7 \pm 5.2 $  & $  138.2 \pm 6.6 $ &   $ -0.3  $ & $   9.5 $ \\  
  54191.9366  &   0.3256 & $ -100.9 \pm 4.8 $  & $  126.6 \pm 6.1 $ &   $ -8.3  $ & $  -1.4 $ \\  
  54217.8096  &   0.6584 & $  126.1 \pm 5.2 $  & $  -84.3 \pm 6.6 $ &   $ -1.2  $ & $  -3.5 $ \\  
  54224.8170  &   0.8111 & $  142.7 \pm 4.9 $  & $  -90.4 \pm 6.2 $ &   $  4.1  $ & $   1.1 $ \\  
  54250.7786  &   0.1963 & $  -98.8 \pm 5.2 $  & $  132.0 \pm 6.6 $ &   $  0.7  $ & $  -2.6 $ \\  
  54514.0334  &   0.2056 & $  -99.5 \pm 5.2 $  & $  139.2 \pm 6.6 $ &   $  2.3  $ & $   2.4 $ \\  
  54520.9910  &   0.3288 & $  -88.4 \pm 4.8 $  & $  123.2 \pm 6.1 $ &   $  3.0  $ & $  -3.7 $ \\  
  54546.9391  &   0.7061 & $  142.6 \pm 5.0 $  & $ -102.2 \pm 6.4 $ &   $ -0.5  $ & $  -6.6 $ \\  
  54574.9033  &   0.2781 & $  -99.5 \pm 4.7 $  & $  145.1 \pm 6.0 $ &   $  5.2  $ & $   5.6 $ \\  
  54578.8906  &   0.6411 & $  114.6 \pm 4.4 $  & $  -78.2 \pm 5.6 $ &   $ -4.6  $ & $  -5.2 $ \\  
  54579.8440  &   0.2061 & $  -96.5 \pm 4.7 $  & $  140.3 \pm 5.9 $ &   $  5.4  $ & $   3.5 $ \\  
  54602.7922  &   0.8056 & $  142.7 \pm 4.9 $  & $  -85.6 \pm 6.2 $ &   $  2.6  $ & $   7.3 $ \\  
  54633.7734  &   0.1656 & $  -91.6 \pm 5.2 $  & $  124.1 \pm 6.6 $ &   $ -2.4  $ & $  -0.6 $ 
\enddata
\tablenotetext{a}{Computed with the ephemeris in Sect.~\ref{sec:ephemeris}.}
\end{deluxetable*}

\begin{figure}
\epsscale{1.15}
\plotone{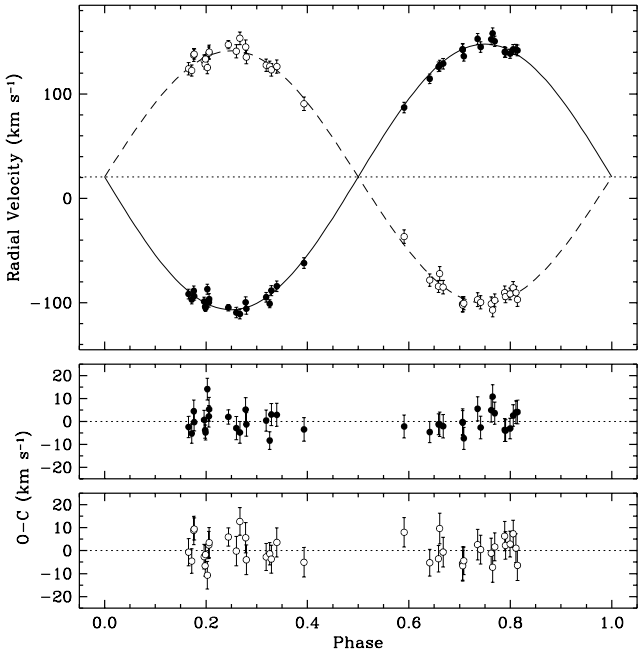}
\caption{\emph{Top:} Measured radial velocities for \aq\ along with
our best-fit orbit model. Solid circles correspond to the photometric
primary (hotter and less massive star), and open circles to the
secondary. The horizontal dotted line represents the center-of-mass
velocity. Phase 0.0 corresponds to the eclipse of the hotter
component. \emph{Bottom:} $O-C$ residuals from the best orbital fit
(same symbols as above).\label{fig:sborbit}}
\end{figure}

A spectroscopic orbit was derived from these measurements with the
orbital period and epoch of the photometric primary eclipse held fixed
at their values determined in Sect.~\ref{sec:ephemeris}. Fits allowing
for a non-zero eccentricity resulted in a value not significantly
different from zero. Consequently for the final solution we adopted a
circular orbit. The elements we obtained are listed in
Table~\ref{tab:sborbit}, and the observations along with this fit are
displayed in Figure~\ref{fig:sborbit}.

\begin{deluxetable}{lc}
\tablewidth{0pc}
\tablecaption{Spectroscopic orbital solution for \aq.\label{tab:sborbit}}
\tablehead{
\colhead{
\hfil~~~~~~~~~~~~~Parameter~~~~~~~~~~~~~~} & \colhead{Value}}
\startdata
\multicolumn{2}{l}{Orbital elements\hfil}                                 \\
~~~~$P$ (days)\tablenotemark{a}                &  1.68743059                 \\
~~~~${\rm Min~I}$ (HJD$-$2,400,000)\tablenotemark{a}             & 53,399.982270  \\
~~~~$\gamma$ (\kms)                            & +20.58~$\pm$~0.58\phs\phn   \\
~~~~$K_{\rm A}$ (\kms)                         &   127.27~$\pm$~0.80\phn\phn \\
~~~~$K_{\rm B}$ (\kms)                         &   120.8~$\pm$~1.0\phn\phn \\
~~~~$e$                                        & 0.0 (fixed)                 \\
\multicolumn{2}{l}{Derived quantities\hfil} \\
~~~~$M_{\rm A}\sin^3 i$ ($M_{\sun}$)\tablenotemark{b}           &  1.300~$\pm$~0.023          \\
~~~~$M_{\rm B}\sin^3 i$ ($M_{\sun}$)\tablenotemark{b}           &  1.369~$\pm$~0.021          \\
~~~~$q\equiv M_{\rm B}/M_{\rm A}$              &  1.054~$\pm$~0.011          \\
~~~~$a_{\rm A}\sin i$ (10$^6$ km)              &  2.953~$\pm$~0.018          \\
~~~~$a_{\rm B}\sin i$ (10$^6$ km)              &  2.803~$\pm$~0.023          \\
~~~~$a \sin i$ ($R_{\sun}$)\tablenotemark{b}                    &  8.274~$\pm$~0.043          \\
\multicolumn{2}{l}{Other quantities pertaining to the fit\hfil}              \\
~~~~$N_{\rm obs}$                              & 39                          \\
~~~~Time span (days)                           &  1559.9                     \\
~~~~$\sigma_{\rm A}$ (\kms)                    & 4.6                         \\
~~~~$\sigma_{\rm B}$ (\kms)                    & 5.9                          
\enddata
\tablenotetext{a}{Ephemeris adopted from Sect.~\ref{sec:ephemeris}.}
\tablenotetext{b}{Based on the physical constants $GM_{\sun}$ and $R_{\sun}$ adopted by \cite{Torres:10a}.}
\end{deluxetable}

\section{Ephemeris}
\label{sec:ephemeris}

A total of 48 times of eclipse for \aq\ were gathered from the
literature\footnote{\texttt{http://www.bav-astro.de/LkDB/index.php?lang=en}},
and were obtained by photographic, visual, photoelectric, or CCD
techniques. From the 27 photoelectric/CCD measurements we determined a
preliminary ephemeris, and detected no significant trends indicative
of any period changes in the $O\!-\!C$ residuals. A large number of
additional timing measurements (167 in total) were derived from our
own URSA and NFO differential photometry described previously. Of
these, 39 timings are based on eclipse events with reasonably good
coverage, i.e., with observations on both the ascending and descending
branches of the primary or secondary minima. These were measured using
either the traditional \cite{Kwee:56} method (KvW) or a parabolic fit,
or with an alternate technique relying on fitting a synthetic
light-curve model to the observations, with the model being computed
using the Wilson-Devinney code \citep[WD;][]{Wilson:71} and subsequent
improvements by \cite{Vaz:07} pertaining to the ephemeris
determination. For the latter method we held all light-curve
parameters fixed to values close to our final solutions reported
later, and adjusted only the time of eclipse. The results from these
three procedures were then weight-averaged (see below). We considered
measurements from URSA and NFO separately. For the 128 URSA/NFO
eclipses with only partial coverage, many having observations on only
one of the branches, we first predicted the approximate center of the
event using the preliminary ephemeris above, and then adjusted this
value using the WD modeling just described.

Realistic uncertainties for these eclipse timings are not easy to
establish, and can depend not only on the quality of the measurements,
but also in our case on the method used to determine them. We
proceeded as follows. We initially considered the uncertainties from
our own measurements to be equal to the internal errors from each
method, and solved for a linear ephemeris adjusting (scaling) these
uncertainties by iterations so as to achieve reduced $\chi^2$ values
near unity. This was done separately by method (KvW, parabolic, or WD
fits), telescope (URSA, NFO), and binary component (primary,
secondary), twelve groups in all. Similarly, for the measurements from
the literature we considered the photographic and visual timings
together as a group, and the CCD and photoelectric timings as another,
separately for the primary and secondary.  For the final fit, minima
measured from our URSA or NFO data by more than one method were merged
together into weighted averages with corresponding uncertainties.  All
215 timings (124 for the primary eclipse, 91 for the secondary) are
reported in Table~\ref{tab:minima} along with their final, rescaled
errors. The resulting linear ephemeris (HJD) is
\begin{equation}
{\rm Min~I} = 2,\!453,\!399.982270(47) + 1.68743059(17) E~,
\label{eq:ephemeris}
\end{equation}
with the figures in parentheses representing uncertainties in units of
the last significant digit.  Residuals from the above fit are listed
in Table~\ref{tab:minima}, and show no obvious pattern as a function
of time. Using only the secondary timings we find a mean phase for the
secondary eclipse of $0.50010 \pm 0.00008$. This is consistent with
0.5, supporting our assumption of a circular orbit in our analysis
below.

\begin{deluxetable}{lccr}
\tablecaption{Eclipse timing measurements for \aq.\label{tab:minima}}
\tablewidth{220pt}
\tablehead{
\colhead{HJD\tablenotemark{a}} &
\colhead{Type\tablenotemark{b}} &
\colhead{Source\tablenotemark{c}} &
\colhead{$O\!-\!C$\tablenotemark{d}}
\\
\colhead{($2,\!400,\!000+$)} &
\colhead{} &
\colhead{} &
\colhead{(days)}
}
\startdata
  28333.220(35)       &   p & P & $ 0.0192$ \\
  28371.154(19)       &   s & P & $-0.0140$ \\
  28655.561(35)       &   p & P & $ 0.0609$ \\
  28661.408(19)       &   s & P & $ 0.0019$ \\
  28679.129(105)      &   p & P & $ 0.0049$ 
\enddata
\tablecomments{Table~\ref{tab:minima} is published in its entirety in
  the electronic edition of the journal. A portion is shown here for
  guidance regarding its form and content.}
\tablenotetext{a}{Uncertainties are given in parenthesis in units of
the last significant digit, and include the scaling described in the
text.}
\tablenotetext{b}{The eclipse type is `p' for primary (hotter and less
massive star, behind during the deeper minimum), and `s' for
secondary.}
\tablenotetext{c}{Source of the measurement: `P' (photographic), `V'
(visual), and `C' (CCD or photoelectric) are for determinations taken
from the literature; `U' or `N' are for new measurements from
incompletely covered minima gathered with URSA or NFO, respectively,
and fit using a WD model (see text); `u' or `n' are for weighted means
of new determinations made with the \citet{Kwee:56} method, a
parabolic fit, and the WD procedure, for minima with observations on
both the descending and ascending branches.}
\tablenotetext{d}{Residuals are based on the ephemeris of
Eq.\,(\ref{eq:ephemeris}). The standard deviation of the (weighted)
residuals is 0.0042 days for the primary timings, 0.0050 days for the
secondary timings, and 0.0046 days for all residuals combined.}
\end{deluxetable}

\section{Light curve solutions}
\label{sec:LCanalysis}

The light curves of \aq\ show moderate proximity effects, with the
curvature between the minima being mostly due to the deformation of
the components and, to a smaller degree, to the mutual
illumination. The small but significant difference in depth between
the primary and secondary eclipses indicates a slightly cooler
temperature for the secondary star, which in this case corresponds to
the more massive and presumably more evolved component.

The analysis of the differential photometry of \aq\ was carried out
using a version of the WD model \citep{Wilson:71, Wilson:79,
Wilson:93} extensively improved as described by \citet{Vaz:07} and
references therein. The URSA and NFO light curves were modeled both
separately and together, adopting the ephemeris in
Eq.(\ref{eq:ephemeris}). The orbit was assumed to be circular, based
on the evidence from the eclipse timings presented above and from the
spectroscopic analysis.  The main quantities we adjusted are the
orbital inclination angle, $i$, the temperature of the secondary,
$T_{\rm eff}^{\rm B}$, the gravitational pseudo-potentials, $\Omega$,
an arbitrary phase shift, $\Delta\phi$, and a luminosity normalization
factor. The primary temperature was held fixed at the value $T_{\rm
eff}^{\rm A} = 6430$\,K described in Sect.~\ref{sec:absdim}, and the
mass ratio was fixed at the value listed in Table~\ref{tab:sborbit}.
Because the orbital period is short, we assumed both components have
their rotation synchronized with the orbital motion. For the
bolometric reflection albedos, $A$, we explored two different
treatments: in one we held them fixed at the value of 0.5 appropriate
for stars with convective envelopes such as these, and in the other we
allowed them to vary freely as the iterations proceeded. The
gravity-brightening exponents $\beta$ were computed using the local
value of $T_{\rm eff}$ for each point on the stellar surfaces, taking
into account mutual illumination following \cite{Alencar:97} and
\cite{Alencar:99}. The radiated flux of both components was described
using the PHOENIX atmosphere models \citep{allardhauschildt1995,
allardetal1997, hauschildtetal1997a, hauschildtetal1997b}. The
luminosity of the secondary was calculated internally from its size
and $T_{\rm eff}$.  All solutions were performed by alternating
between the least-squares and simplex methods to improve convergence
\citep[see, e.g.,][]{Press:92}, with equal weights for all
measurements. Iterations were stopped when the corrections to the
individual elements were at least one order of magnitude smaller than
the formal errors, and they oscillated between positive and negative
in consecutive iterations.

\begin{deluxetable*}{rrrrrrrl}
\tablecaption{Light-curve solutions for \aq\ to explore different
limb-darkening laws and different treatments of the reflection
albedos.\label{tab:LD}}
\tablewidth{350pt}
\tablehead{
\colhead{$i~(^{\circ})$} & 
\colhead{$T_{\rm eff}^{\rm{B}}$ (K)} & 
\colhead{$A_{\rm A}$} & 
\colhead{$A_{\rm B}$} &
\colhead{$\Omega_{\rm A}$} & 
\colhead{$\Omega_{\rm B}$} &
\colhead{$\Delta\phi$} &
\colhead{${\sigma {\rm (mmag)}}\atop{\scriptscriptstyle \mathrm{URSA\,{\displaystyle /}\,NFO}}$}
}
\startdata
\multicolumn{8}{c}{LINEAR limb darkening law}\\[1pt]
  81.2553  &  6347.78  &  0.5000   &  0.5000   & 4.7765   &  4.6224   &   1.18     &  9.015\,/\,5.487 \\
  $\pm 52$ &  $\pm 66$ &  fixed    &  fixed    & $\pm 14$ &  $\pm 11$ &   $\pm 10$ &                  \\[+2pt]

  81.3235  &  6343.17  &  0.4216   &  0.3972   & 4.7777   &  4.6358   &   1.21     &  9.010\,/\,5.460 \\
  $\pm 76$ &  $\pm 72$ &  $\pm 79$ &  $\pm 73$ & $\pm 16$ &  $\pm 13$ &   $\pm 12$ &                  \\
\noalign{\smallskip\hrule\smallskip}
\multicolumn{8}{c}{LOGARITHMIC limb darkening law}\\[1pt]
  81.3498  &  6346.82  &  0.5000   &  0.5000   & 4.7659   &  4.6307   &   1.15     &  9.034\,/\,5.507 \\
  $\pm 51$ &  $\pm 66$ &  fixed    &  fixed    & $\pm 14$ &  $\pm 12$ &   $\pm 10$ &                  \\[+2pt]

  81.4415  &  6341.77  &  0.4028   &  0.3762   & 4.7664   &  4.6496   &   1.19     &  9.016\,/\,5.482 \\
  $\pm 77$ &  $\pm 71$ &  $\pm 82$ &  $\pm 75$ & $\pm 15$ &  $\pm 15$ &   $\pm 10$ &                  \\
\noalign{\smallskip\hrule\smallskip}
\multicolumn{8}{c}{SQUARE ROOT limb darkening law}\\[1pt]
  81.3362  &  6342.54  &  0.5000   &  0.5000   & 4.7665   &  4.6291   &   1.14     &  9.033\,/\,5.508 \\
  $\pm 51$ &  $\pm 66$ &  fixed    &  fixed    & $\pm 14$ &  $\pm 12$ &   $\pm 10$ &                  \\[+2pt]

  81.4319  &  6337.35  &  0.4006   &  0.3722   & 4.7673   &  4.6482   &   1.20     &  9.017\,/\,5.477 \\
  $\pm 79$ &  $\pm 71$ &  $\pm 82$ &  $\pm 74$ & $\pm 16$ &  $\pm 14$ &   $\pm 10$ &                  \\
\noalign{\smallskip\hrule\smallskip}
\multicolumn{8}{c}{QUADRATIC limb darkening law}\\[1pt]
  81.3352  &  6341.43  &  0.5000   &  0.5000   & 4.7663   &  4.6301   &   1.15     &  9.026\,/\,5.498 \\
  $\pm 51$ &  $\pm 66$ &  fixed    &  fixed    & $\pm 14$ &  $\pm 12$ &   $\pm 10$ &                  \\[+2pt]

  81.4231  &  6336.75  &  0.4119   &  0.3855   & 4.7671   &  4.6479   &   1.20     &  9.016\,/\,5.474 \\
  $\pm 79$ &  $\pm 71$ &  $\pm 81$ &  $\pm 74$ & $\pm 16$ &  $\pm 15$ &   $\pm 10$ &                  
\enddata
\tablecomments{Test solutions for a fixed mass ratio of $q \equiv
M_{\rm B}/M_{\rm A} = 1.054 \pm 0.011$
(Sect.~\ref{sec:spectroscopy}). Phase shifts $\Delta\phi$ are in units
of $10^{-4} P$. Uncertainties are given in units of the last
significant digit and represent internal errors from the WD code.}
\end{deluxetable*}

Four different limb-darkening laws were investigated: linear,
quadratic, square-root, and logarithmic \citep[see][]{Claret:00}. The
coefficients for these laws were interpolated from the tables by the
above author using a bilinear scheme for the current values of $T_{\rm
eff}$ and $\log g$ at each iteration. The results of our initial
exploration of the limb-darkening laws are presented in
Table~\ref{tab:LD}, in which we used the URSA and NFO data
simultaneously. These tests were run both holding the reflection
albedos fixed, and allowing them to vary.

\begin{figure}
\epsscale{1.15}
\plotone{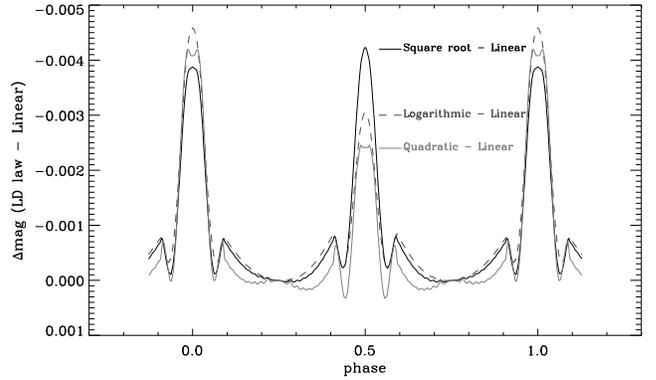}
\figcaption{Differences between light curve models that use a
non-linear limb darkening law and one that adopts the linear law. The
linear law results in brighter stars (deeper minima). The logarithmic
and quadratic laws are the most similar to each other. The maximum
amplitude of the differences compared to the linear law is only
$\sim$5\,mmag.
\label{fig:LD}}
\end{figure}

Despite the larger freedom of the model when using non-linear
limb-darkening laws, we found that there is relatively little
difference in the quality of the fits, and that the linear law gives
marginally better solutions.  Figure~\ref{fig:LD} illustrates the
effect that the non-linear limb-darkening laws have on the light
curves, relative to the effect of the linear law. The maximum
difference for a system such as \aq\ turns out to be quite small
($\sim$5\,mmag), which explains why the resulting light elements in
Table~\ref{tab:LD} are rather similar for the various laws. We also
note a modest improvement in the solutions when adjusting the
bolometric albedos, as opposed to leaving them fixed. The resulting
values of $\beta$ are somewhat smaller than the canonical values.
Based on these tests, for the final solutions we chose to adopt the
linear limb-darkening law and to allow the albedos to be adjusted.

\begin{figure}
\epsscale{1.15}
\plotone{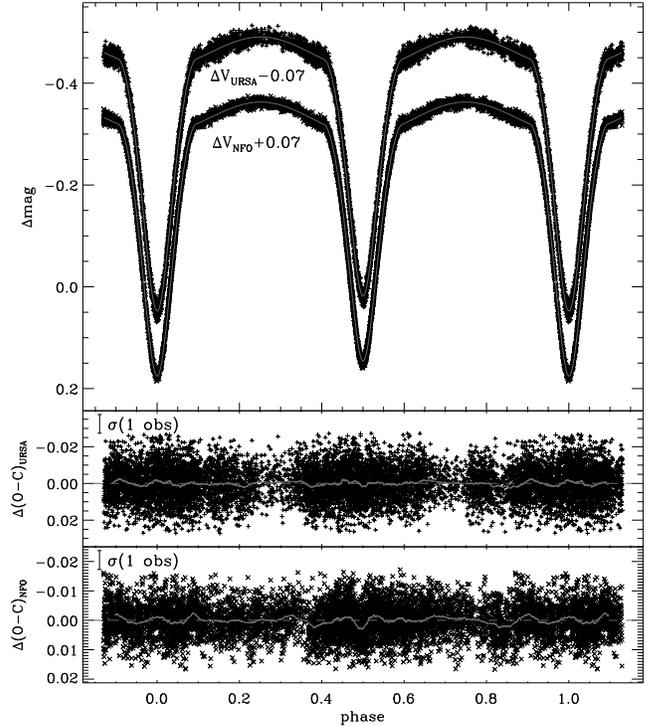}
\figcaption[]{
Differential photometry of \aq\ from the URSA and NFO telescopes,
along with the fitted model. Phase 0.0 corresponds to the eclipse of
the hotter, smaller, and less massive component. The light curves are
shifted vertically for clarity. The lower panels show the $O\!-\!C$
residuals, with the typical error of a single measurement indicated in
the top left corner of each panel. Remaining systematic effects in the
fits are very small, as shown by a boxcar smoothing of the residuals
represented in the lower panels with gray lines. The box size adopted
was 220 points for URSA and 180 points for NFO.\label{fig:lc}}
\end{figure}

Our final results are presented in Table~\ref{tab:wdfinal}, for the
individual solutions to the URSA and NFO data and also for the
combined fit. The uncertainties for the individual solutions are the
internal errors reported by the WD code. For the combined fit that we
adopt for the remainder of the analysis we have conservatively
increased the internal errors by adding in quadrature half of the
difference between the parameters for the individual solutions. Also
included in Table~\ref{tab:wdfinal} are the ``volume'' radii $r_{\rm
vol,A}$ and $r_{\rm vol,B}$, which are used in the next section to
compute the absolute radii of the components.

\begin{deluxetable*}{lrrr|lrrr}
\tablecaption{Final light curve solutions for \aq.\label{tab:wdfinal}}
\tablewidth{420pt}
\tablehead{
\colhead{Parameter} & 
\colhead{URSA} & 
\colhead{NFO} & 
\colhead{Combined}  &
\colhead{Parameter} & 
\colhead{URSA} & 
\colhead{NFO} & 
\colhead{Combined}
}
\startdata
$i$ ($\degr$)                 &  81.282   &  81.3418  &  81.323   & $T_{\rm eff}^{\rm B}$ (K)     &  6344.5   &  6342.46  &  6343.2   \\
                              & $\pm 12$  & $\pm 91$  & $\pm 31$  &                               & $\pm 1.2$ & $\pm 97$  & $\pm 1.2$ \\ [+2pt]
$A_{\rm A}$                   &  0.447    &  0.4142   &  0.422    & $A_{\rm B}$                   &  0.427    &  0.3867   &  0.397    \\
                              & $\pm 14$  & $\pm 97$  & $\pm 81$  &                               & $\pm 13$  & $\pm 88$  & $\pm 21$  \\ [+2pt]
$\Omega_{\rm A}$              &  4.7781   &  4.7783   &  4.7777   & $\Omega_{\rm B}$              &  4.6282   &  4.6389   &  4.6358   \\
                              & $\pm 25$  & $\pm 20$  & $\pm 13$  &                               & $\pm 20$  & $\pm 17$  & $\pm 58$  \\ [+2pt]
$r_{\rm A,pole}$              &  0.26599  &  0.26598  & 0.26602   & $r_{\rm B,pole}$              &  0.2854   &  0.2846   & 0.2848    \\
                              &  $\pm 91$ &  $\pm 88$ & $\pm 85$  &                               &  $\pm 27$ &  $\pm 26$ & $\pm 26$  \\ [+2pt]
$r_{\rm A,point}$             &  0.2839   &  0.2838   & 0.2839    & $r_{\rm B,point}$             &  0.3082   &  0.3070   & 0.3073    \\
                              &  $\pm 10$ &  $\pm 10$ & $\pm 10$  &                               &  $\pm 14$ &  $\pm 14$ & $\pm 15$  \\ [+2pt]
$r_{\rm A,side}$              &  0.2714   &  0.27134  & 0.27138   & $r_{\rm B,side}$              &  0.2922   &  0.2913   & 0.2915    \\
                              &  $\pm 10$ & $\pm 98$  & $\pm 95$  &                               &  $\pm 29$ & $\pm 28$  & $\pm 29$  \\ [+2pt]
$r_{\rm A,back}$              &  0.2794   &  0.2793   & 0.2794    & $r_{\rm B,back}$              &  0.3019   &  0.3008   & 0.3011    \\
                              &  $\pm 12$ & $\pm 12$  & $\pm 12$  &                               &  $\pm 32$ & $\pm 31$  & $\pm 31$  \\ [+2pt]
$r_{\rm A,vol}$               &  0.2725   &  0.2725   & 0.27251   & $r_{\rm B,vol}$               &  0.2934   &  0.2925   & 0.2928    \\
                              & $\pm 11 $ & $\pm 10$  & $\pm 99$  &                               & $\pm 29$  & $\pm 29$  & $\pm 29$  \\ [+2pt]
$\Delta\phi$ ($10^{-4} P$)    &  1.41     &  1.10     &   1.21  & $\ell_{\rm B}/\ell_{\rm A} (V)$ &  1.094    &  1.086    &  1.088    \\
                              & $\pm 18$  & $\pm 13$  & $\pm 10$  &                               & \nodata   & \nodata   &  \nodata  \\ [+2pt]
$\beta_{\rm A}$               &    0.312  &    0.312  & 0.312     & $\beta_{\rm B}$               &    0.320  &    0.320  & 0.320     \\
$x_{\rm bolo,A}$              &  0.3887   &  0.3887   &  0.3887   & $x_{\rm bolo,B}$              &  0.3916   &  0.3916   &  0.3916   \\
$x_{\rm V,A}$ (URSA)          &  0.6748   & \nodata   & 0.6748    & $x_{\rm V,B}$ (URSA)          &  0.6772   & \nodata   & 0.6773    \\
$x_{\rm V,A}$ (NFO)           & \nodata   &  0.6748   & 0.6748    & $x_{\rm V,B}$ (NFO)           & \nodata   &  0.6774   & 0.6773    \\
$\sigma_{\rm URSA}$ (mmag)    &  9.006    & \nodata   & 9.010     & $\sigma_{\rm NFO}$ (mmag)     &  \nodata  & 5.458     & 5.460   
\enddata
\tablecomments{The mass ratio was held fixed at its spectroscopically
determined value of $q \equiv M_{\rm B}/M_{\rm A} = 1.054 \pm
0.011$. The uncertainties in the relative radii $r$ account for those
in the pseudo-potentials and in the mass ratio. The $V$-band light
ratio $\ell_{\rm B}/\ell_{\rm A}$ is calculated at phase 0.25. The
bolometric and passband-specific limb darkening coefficients $x$ were
adjusted during the iterations to follow the evolution of $T_{\rm
eff}$ and $\log g$. The gravity-brightening coefficients $\beta$
varied over the mutually illuminated stellar surfaces, following
\cite{Alencar:97} and \cite{Alencar:99}, and the values reported above
are those for the non-illuminated hemispheres. Uncertainties in the
last column include a contribution from the difference between the
URSA and NFO solutions.}
\end{deluxetable*}

The observations along with the fitted model are shown in
Figure~\ref{fig:lc}, and residuals are displayed in the lower
panels. The remaining systematic effects in the light curve are very
small, as illustrated by the gray curves in the lower panels
representing a running mean of the residuals.  There is good agreement
between the $V$-band light ratio from our final combined fit
($\ell_{\rm B}/\ell_{\rm A} = 1.088$) and the spectroscopic value of
$\ell_{\rm B}/\ell_{\rm A} = 1.05 \pm 0.04$ that we reported in
Sect.~\ref{sec:spectroscopy}, which is in a passband similar to $V$.
This supports the accuracy of our solution, and in particular that of
the relative radii.

\section{Absolute dimensions}
\label{sec:absdim}

Our spectroscopic and photometric analyses lead to the absolute masses
and radii for \aq\ reported in Table~\ref{tab:absdim} below, which
have relative uncertainties smaller than 1.8\% and 1.1\%,
respectively.  Given the short orbital period of the system we have
assumed that each star's rotation is synchronized with the orbital
motion. Our measured $v \sin i$ values from
Sect.~\ref{sec:spectroscopy} are indeed consistent with the expected
synchronous rotational velocities listed in the table, although they
do have fairly large uncertainties.

No spectroscopic determination of the metallicity is available for
\aq.  A rough photometric estimate was derived by means of the
calibration for F stars by \cite{Crawford:75} along with the $uvby$
observations in the Str\"omgren system by \cite{Hilditch:75}, which,
however, lack the necessary measurement of the reddening-free index
$\beta$. We circumvented this by using an estimate of the interstellar
reddening from dust maps following \cite{Hakkila:97},
\cite{Schlegel:98}, and \cite{Drimmel:03}, and an approximate distance
of 580\,pc (see below). These three sources give $E(B-V)$ values of
0.011, 0.039, and 0.036~mag, which are insensitive to distance changes
of $\pm$100\,pc.  We adopt the straight average of $E(B-V) = 0.029 \pm
0.010$. With this value and the $uvby$ photometry the metallicity
inferred for \aq\ is ${\rm [Fe/H]} \approx -0.19$.

As indicated in Sect.~\ref{sec:spectroscopy}, the severe line
broadening does not permit us to obtain reliable spectroscopic
estimates of the effective temperatures of the components. A mean
temperature for the system may be derived from standard photometry
available in the literature, including $JH\!K_s$ measurements from
2MASS \citep{Cutri:03}, $V_{\rm T}$ and $B_{\rm T}$ from the Tycho-2
catalog \citep{Hog:00}, $V$ and Str\"omgren $b-y$ as reported by
\cite{Hilditch:75}, Johnson $B$ and $V$ from the APASS catalog
\citep{Henden:12}, and Johnson-Cousins $VI_{\rm C}$ photometry from
the TASS catalog \citep{Droege:06}. A total of eleven, non-independent
color indices were formed for which color/temperature calibrations
have been established by \cite{Casagrande:10}. Appropriate reddening
corrections for each of the indices were applied following
\cite{Cardelli:89}, using the $E(B-V)$ value established above. The
resulting temperatures for a metallicity of ${\rm [Fe/H]} = -0.19$ are
given in Table~\ref{tab:teff}, and their weighted average is $T_{\rm
eff} = 6380 \pm 40$\,K.  We adopt a more conservative error for this
analysis of 100\,K. The metallicity dependence of the average
temperature is very small: assuming solar metallicity would lower it
by only 16\,K. Based on this mean photometric temperature for the
system and a preliminary temperature ratio from our light curve
solutions, we inferred a temperature for the hotter star of $T_{\rm
eff}^{\rm A} = 6430 \pm 100$\,K, corresponding to spectral type F5.
This is the value employed in our final fits described in
Sect.~\ref{sec:LCanalysis}. The temperature derived for the cooler
star from our solutions is $T_{\rm eff}^{\rm B} = 6340 \pm 100$\,K,
which corresponds approximately to an F6 star. The two temperatures
are of course highly correlated with each other, and the temperature
\emph{difference} is much better determined than the absolute values,
as it is directly related to the well-measured difference in eclipse
depths. We estimate the difference as $\Delta T_{\rm eff} = 90 \pm
20$\,K.

\begin{deluxetable}{lccc}
\tablewidth{0pc}
\tablecaption{Color indices and mean effective temperatures for \aq.\label{tab:teff}}
\tablehead{
\colhead{Index} & \colhead{Value (mag)} & \colhead{$T_{\rm eff}$ (K)} & Source
}
\startdata
 Johnson $B-V$                 &  $0.481 \pm 0.057$  &  $6388 \pm 241$     &   1 \\
 Johnson-Cousins $V-I_{\rm C}$ &  $0.632 \pm 0.215$  &  $6153 \pm 542$     &   2 \\
 2MASS $V-J$                   &  $0.949 \pm 0.026$  &  $6475 \pm 103$     &   3\,,\,4 \\
 2MASS $V-H$                   &  $1.176 \pm 0.025$  &  $6392 \pm \phn77$  &   3\,,\,4 \\
 2MASS $V-K_s$                 &  $1.235 \pm 0.021$  &  $6426 \pm \phn71$  &   3\,,\,4 \\
 2MASS $J-K_s$                 &  $0.286 \pm 0.031$  &  $6239 \pm 210$     &   4 \\
 Tycho $B_{\rm T}-V_{\rm T}$   &  $0.556 \pm 0.081$  &  $6288 \pm 267$     &   5 \\
 Tycho-2MASS $V_{\rm T}-J$     &  $1.084 \pm 0.062$  &  $6296 \pm 169$     &   4\,,\,5 \\
 Tycho-2MASS $V_{\rm T}-H$     &  $1.311 \pm 0.061$  &  $6264 \pm 118$     &   4\,,\,5 \\
 Tycho-2MASS $V_{\rm T}-K_s$   &  $1.370 \pm 0.060$  &  $6307 \pm 120$     &   4\,,\,5 \\
 Str\"omgren $b-y$             &  $0.327 \pm 0.030$  &  $6401 \pm 208$     &   3 
\enddata
\tablecomments{
Sources are: 
1.~\cite{Henden:12};
2.~\cite{Droege:06};
3.~\cite{Hilditch:75};
4.~\cite{Cutri:03};
5.~\cite{Hog:00}.
Temperature uncertainties include contributions from the photometry,
reddening, and estimated systematic errors as well as the scatter in
the calibrations following \cite{Casagrande:10}. The metallicity
adopted is ${\rm [Fe/H]} = -0.19$. In computing the temperatures all
photometric indices were corrected for reddening following
\cite{Cardelli:89}, using $E(B-V) = 0.029 \pm 0.010$.}
\end{deluxetable}

\begin{deluxetable}{lcc}
\tablewidth{0pt}
\tablecaption{Physical properties of \aq.\label{tab:absdim}}
\tablehead{
\colhead{Parameter} &
\colhead{Star A} & 
\colhead{Star B}
}
\startdata
\multicolumn{3}{l}{Absolute dimensions} \\ [+1ex]
~~~Mass ($M_{\sun}$)             &  1.346~$\pm$~0.024       &  1.417~$\pm$~0.022    \\
~~~Radius ($R_{\sun}$)           &  2.281~$\pm$~0.014       &  2.451~$\pm$~0.027    \\
~~~$\log g$ (cgs)                & 3.8504~$\pm$~0.0094      &  3.810~$\pm$~0.012    \\
~~~$v_{\rm sync} \sin i$ (\kms)  &   67.6~$\pm$~0.4\phn     &   72.6~$\pm$~0.8\phn  \\
~~~$v \sin i$ (\kms) \tablenotemark{a}    &     59~$\pm$~10          &     73~$\pm$~10       \\
~~~$a$ ($R_{\sun}$)              &      \multicolumn{2}{c}{8.370~$\pm$~0.044}       \\ [+1ex]
\multicolumn{3}{l}{Radiative and other properties} \\ [+1ex]
~~~$T_{\rm eff}$ (K)             &   6430~$\pm$~100\phn     &   6340~$\pm$~100\phn  \\
~~~$\log L/L_\sun$               &  0.901~$\pm$~0.027       &  0.939~$\pm$~0.042    \\
~~~$M_{\rm bol}$ (mag)           &  2.479~$\pm$~0.069       &   2.38~$\pm$~0.10     \\
~~~${\rm BC}_V$ (mag) \tablenotemark{b}            &   0.00~$\pm$~0.10        &  $-$0.01~$\pm$~0.10\phs \\
~~~$M_V$ (mag)                   &   2.48~$\pm$~0.12        &   2.39~$\pm$~0.14     \\
~~~$L_{\rm B}/L_{\rm A}$         &      \multicolumn{2}{c}{1.09~$\pm$~0.13}         \\
~~~$(L_{\rm B}/L_{\rm A})_V$     &      \multicolumn{2}{c}{1.08~$\pm$~0.19}         \\
~~~$E(B-V)$ (mag)                &      \multicolumn{2}{c}{0.029~$\pm$~0.010}       \\
~~~Distance (pc)                 &      \multicolumn{2}{c}{577~$\pm$~27\phn}    
\enddata
\tablecomments{Star A (photometric primary) corresponds to the hotter
and less massive star of the pair.}
\tablenotetext{a}{Values measured spectroscopically.}
\tablenotetext{b}{Bolometric corrections from \cite{Flower:96}, with conservative uncertainties.}
\end{deluxetable}

Additional quantities listed in Table~\ref{tab:absdim} include the
luminosities and the absolute visual magnitudes, for which we adopted
bolometric corrections from \cite{Flower:96} with conservative
uncertainties of 0.10 mag. Alternate bolometric correction tables such
as those of \cite{Popper:80} or \cite{Schmidt-Kaler:82} give very
similar results when used with consistent bolometric magnitudes for
the Sun \citep[see][]{Torres:10b}. The distance to \aq\ is estimated
to be $577 \pm 27$ pc, based on the combined out-of-eclipse magnitude
of $V = 10.575 \pm 0.010$ \citep{Hilditch:75} and the extinction
computed as $A_V = 3.1 \times E(B-V)$. Separate distances calculated
for the individual components using the measured light ratio agree
with the above value within 1~pc, indicating a high degree of internal
consistency in the parameters.

\section{Comparison with stellar evolution models}
\label{sec:stellarevolution}

The masses of the \aq\ components are both in the regime in which
stars develop convective cores, and offer a valuable opportunity for a
comparison with stellar evolution theory regarding the importance of
extra mixing beyond the core, which has been found to be necessary in
order to reproduce observations of binary stars and star clusters
\cite[see, e.g.,][and references therein]{Andersen:90, Chiosi:92,
Chiosi:99}. This extended mixing can result from turbulent flows
moving across the standard convective boundary defined by the
classical \cite{Schwarzschild:06} criterion, usually referred to as
``overshooting'', or it can also be generated in part by differential
rotation and the associated shear layer that develops at the edge of
the convective core \citep[see, e.g.,][]{Pinsonneault:91}. Here we
will assume that the extra mixing comes only from overshooting from
the core, which is the way the effect is most commonly described in
current models. The prescriptions for this vary from model to model,
and in general the treatment is still very much ad hoc.

\begin{figure}[t!]
\epsscale{1.0}
\plotone{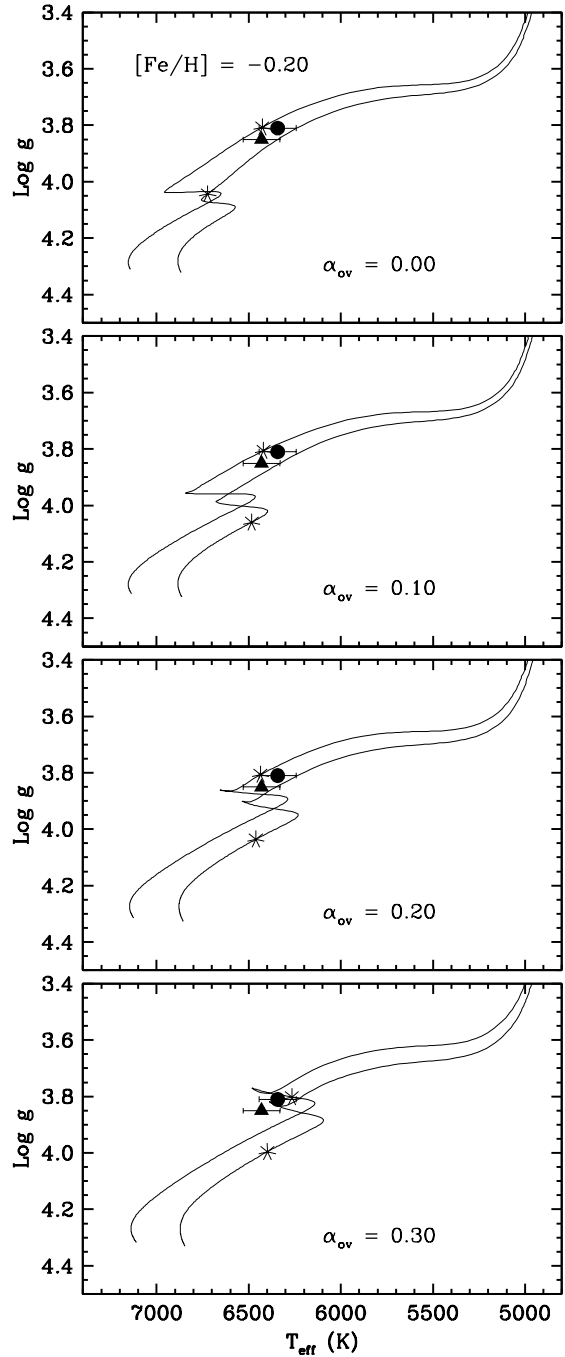}
\figcaption[]{
Evolutionary tracks for the measured masses of the components of \aq\
from the Granada models of \cite{Claret:04}. The more massive star is
represented with a filled circle, and the other with a triangle. The
best-fit metallicity for $\alpha_{\rm ov} = 0.00$ (no overshooting) is
$Z = 0.012$, corresponding to ${\rm [Fe/H]} = -0.20$ \citep[$Z_{\sun}
= 0.0189$;][]{Grevesse:98}. The lower panels show the effect of
increasing $\alpha_{\rm ov}$ at the same metallicity. An asterisk on
the track for the more massive star (left) indicates the best-fit
location, and the asterisk on the other track is the expected location
of the less massive star at the same age as the other. This
illustrates the age discrepancy mentioned in the text.
\label{fig:granada}}
\end{figure}

We begin by considering the Granada evolutionary models of
\cite{Claret:04}, in which the overshoot length is taken to be $d_{\rm
ov} = \alpha_{\rm ov} H_p$, where $H_p$ is the local pressure scale
height at or across the formal edge of the convective core.
Figure~\ref{fig:granada} shows the evolutionary tracks for the
measured masses of \aq, and a range of overshooting parameters
(assumed to be the same for the two stars) from $\alpha_{\rm ov} =
0.00$ (no overshooting) to 0.30.  For models with no overshooting the
best match to the temperatures was found for a metallicity of $Z =
0.012$. This corresponds to ${\rm [Fe/H]} = -0.20$ in these models,
which is very close to our photometric estimate for the
system. However, the fit places both stars in the Hertzprung gap,
which is a very rapid and a priori unlikely state of
evolution. Furthermore, the models predict the more massive star
(filled circle) to be the hotter one, which is the opposite of what we
observe. Better agreement with the measured temperature difference
would require nearly identical masses (to well within 1\%; see below),
while our spectroscopic analysis shows them to differ by about 5\%
(Table~\ref{tab:sborbit}). The measured mass ratio is therefore
inconsistent with a post-main-sequence status for \aq, and this argues
for a significant amount of extra mixing.  Indeed, only when the
overshooting parameter reaches a value near $\alpha_{\rm ov} = 0.30$
is it possible to obtain a better match to the temperature difference
at this metallicity (Figure~\ref{fig:granada}, bottom panel), and this
places the stars at the very end of the main-sequence phase. Even this
fit is unsatisfactory, however, as the models predict different ages
for stars of these masses and radii, with the more massive one
appearing younger.

\begin{figure}[t!]
\epsscale{1.15}
\plotone{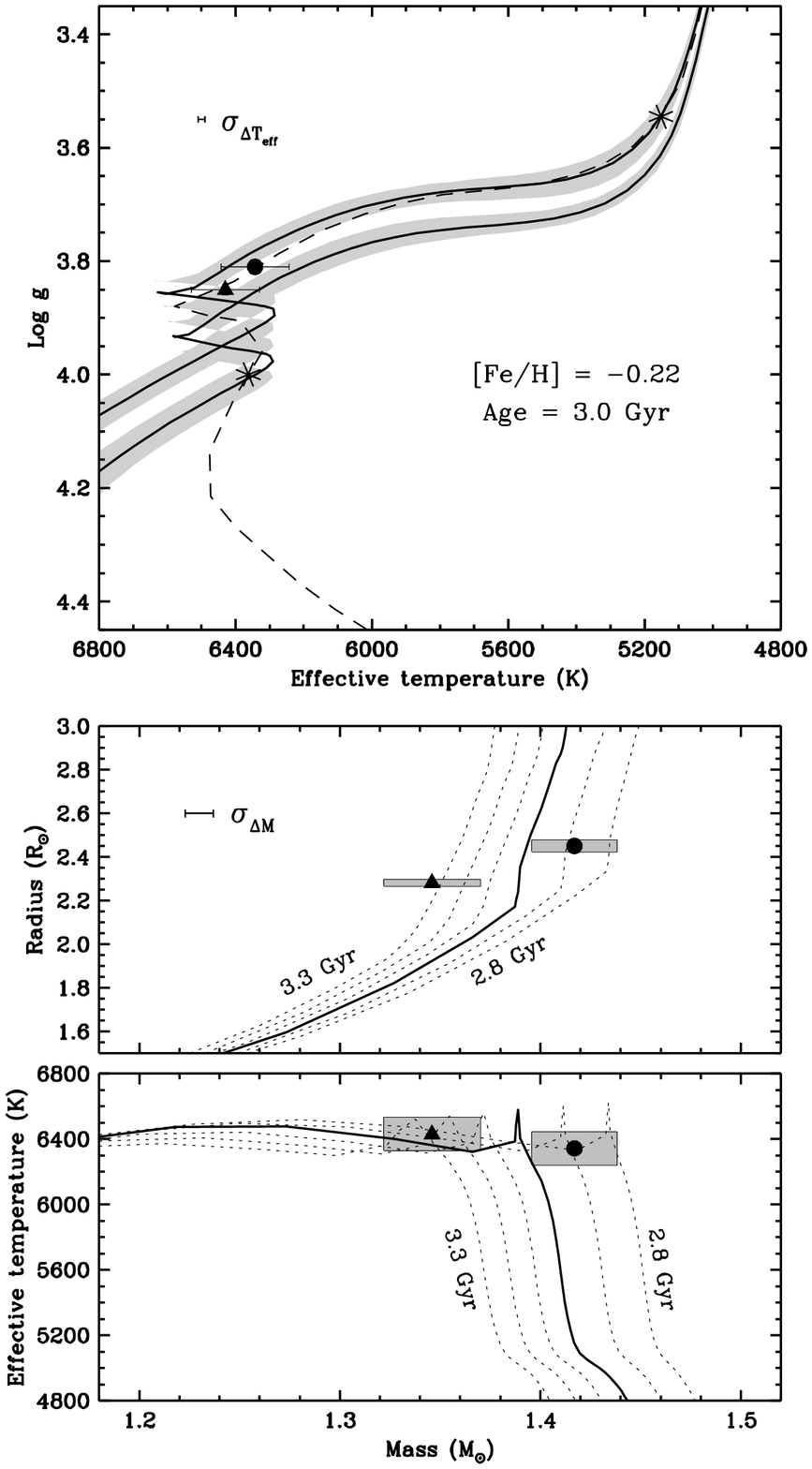}
\figcaption[]{
Measurements for \aq\ compared against the Yonsei-Yale stellar
evolution models \citep{Yi:01, Demarque:04} for $Z = 0.0113$,
corresponding to ${\rm [Fe/H]} = -0.22$ for these calculations
\citep[$Z_{\sun} = 0.0181$;][]{Grevesse:93a}. Interpolations in mass,
metallicity, and age were performed with routines provided by the
modelers. The $\alpha$-element enhancement is assumed to be
[$\alpha$/Fe] = 0.0.  \emph{Top:} Evolutionary tracks for the measured
masses are shown with solid lines, the shaded area indicating the
uncertainty in the location of the tracks stemming from the mass
errors. The filled circle and triangle correspond to the more massive
and less massive star, respectively. The best-fit isochrone for an age
of 3.0~Gyr is shown with a dashed line. The predicted mass ratio for
two evolved stars of this age so near each other in this part of the
diagram is $q = 1.0028$, which is significantly closer to unity than
measured. This is illustrated with the asterisks, which mark the
stars' location on the isochrone for their measured masses.
\emph{Bottom:} Mass/radius and mass/temperature diagrams showing
isochrones in steps of 0.1~Gyr from 2.8 to 3.3~Gyr. The solid line
represents the best-fit isochrone from the top panel, and we indicate
also the uncertainty in the mass difference.
\label{fig:yale}}
\end{figure}

We explored this further by considering other published series of
stellar evolution calculations, although in this case the overshooting
parameter is generally fixed at a value chosen by the modelers and
cannot be changed by the user. In the Yonsei-Yale calculations by
\cite{Yi:01} overshooting is treated in the same way as the Granada
models, and the multiplicative factor $\alpha_{\rm ov}$ ramps up
gradually from zero for stars with no convective core to a maximum of
0.20 as the mass increases \citep[see][]{Demarque:04}, and is a
function of metallicity. For models with solar metallicity the mass
interval over which the overshooting parameter rises from $\alpha_{\rm
ov} = 0.00$ to 0.20 is 1.2--1.4\,$M_{\sun}$. The $\log g$ vs.\ $T_{\rm
eff}$ diagram in the top panel of Figure~\ref{fig:yale} shows the
measurements of \aq\ compared against Yonsei-Yale evolutionary tracks
(solid lines) for the measured masses of the components. The
metallicity in the models has been adjusted to a value of $Z = 0.0113$
(or ${\rm [Fe/H]} = -0.22$, similar to the value inferred using the
Granada models) that gives the best fit with the stars just past the
point of hydrogen exhaustion.  The corresponding $\alpha_{\rm ov}$
values for these masses are approximately 0.20 for the cooler and more
massive star and 0.12 for the other.  The best-fit isochrone has an
age of 3.0~Gyr, and as before, the fit places the stars in the
Hertzprung gap. However, once again there is a problem with the
masses. Evolution is such a strong function of mass at these phases
that in order to be so close together in this part of the diagram the
Yonsei-Yale models require stars of this average mass to have
virtually identical masses ($q = 1.0028$).  This is in strong
disagreement with the measured mass ratio ($q = 1.054 \pm 0.011$) at
the 4.6$\sigma$ level, and excludes such an evolved state. Instead the
stars must still be on the main sequence if they are to have the same
age, and this implies there must be a significant degree of extra
mixing.  We illustrate the mass disagreement in the figure in another
way by marking the expected location of the two stars on the best-fit
isochrone at their measured masses. Theory predicts them to be much
farther apart in $T_{\rm eff}$ and $\log g$ than observed.

An equivalent way of interpreting the discrepancy is in terms of age,
as noted earlier. The lower panels of Figure~\ref{fig:yale} show the
predictions from the Yonsei-Yale models for the radius and temperature
as a function of mass, along with isochrones for a range of ages. The
more massive star is better fit at a younger age than the secondary,
the difference being about 0.45~Gyr (or 15\%), as seen more clearly in
the mass/radius plane.  Isochrones in this diagram are mostly vertical
for stars of this size, so the significance of the age difference
depends largely on the mass separation. The mass uncertainties shown
by the shaded boxes represent \emph{total} errors; the error in the
mass \emph{difference}, $\sigma_{\Delta M}$, is considerably smaller
and is also indicated in the figure. Therefore, the age discrepancy is
highly significant.

\begin{figure}[t!]
\epsscale{1.15}
\plotone{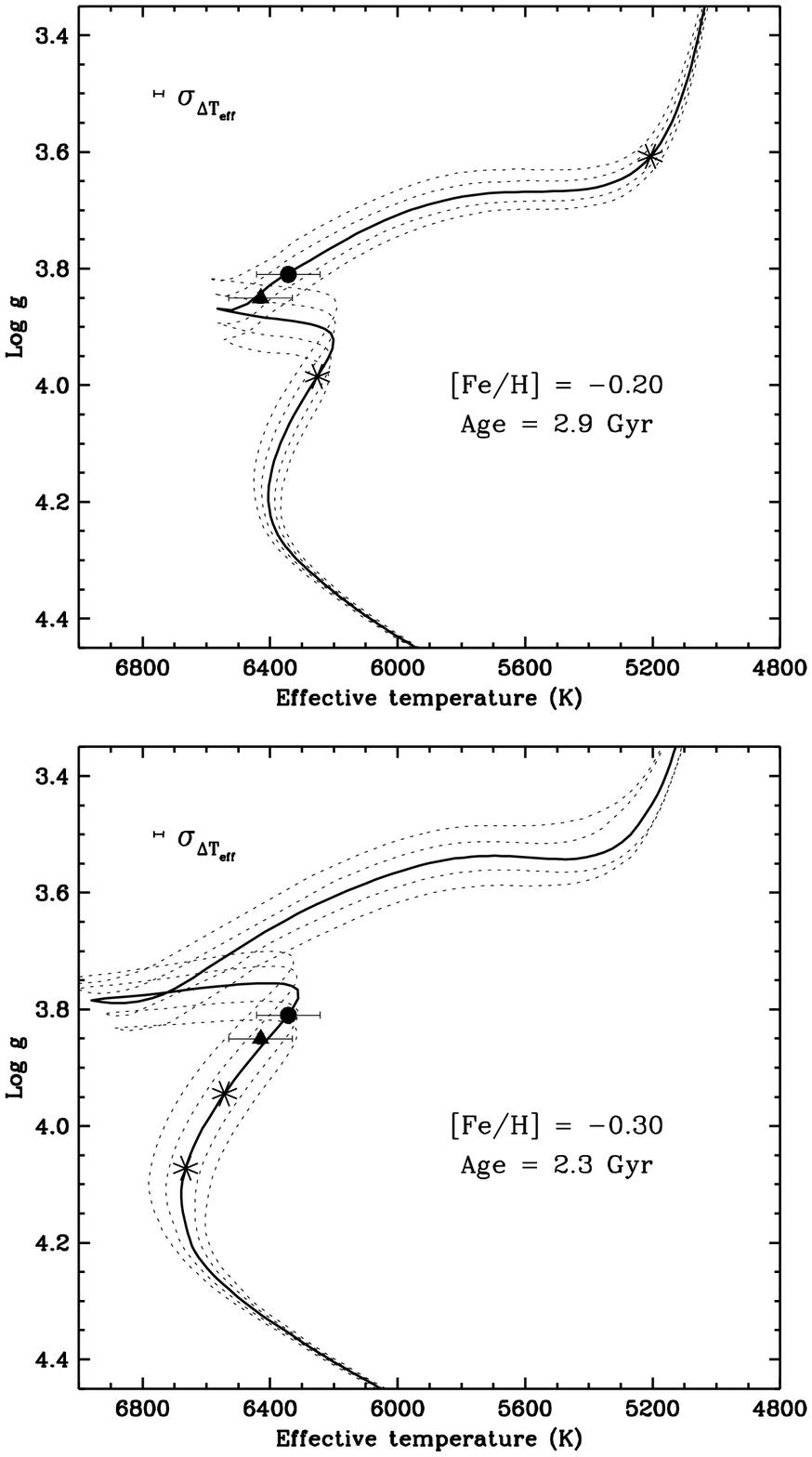}
\figcaption[]{
Similar to the top panel of Figure~\ref{fig:yale}, showing the
comparison of the measurements for \aq\ against the Victoria-Regina
models \citep{VandenBerg:06}. Age interpolation was performed with the
routines provided by the modelers. \emph{Top:} Isochrones are shown
for ages of 2.7 to 3.1~Gyr, with the best-fit model represented with
the solid line for an age of 2.9~Gyr. This fit has $Z = 0.0125$
\citep[or ${\rm [Fe/H]} = -0.20$ with $Z_{\sun} =
0.0188$;][]{Grevesse:93b}, and places the stars beyond the point of
hydrogen exhaustion. The predicted mass ratio matching the measured
radii and temperatures is $q = 1.0008$, much smaller than observed.
\emph{Bottom:} Alternate solution in which the best-fit metallicity
and age ($Z = 0.0100$ or ${\rm [Fe/H]} = -0.30$, and 2.3 Gyr) were
adjusted in such a way as to permit the stars to be in a more likely
evolutionary state just prior to the blue hook. The mass ratio
predicted by theory ($q = 1.0169$) is still significantly smaller than
measured. Isochrones are shown for ages of 2.1 to 2.5 Gyr. In both
panels the asterisks mark the location of the stars on the best-fit
isochrone according to their measured mass.
\label{fig:victoria}}
\end{figure}

As mentioned in Sect.~\ref{sec:absdim}, \aq\ lacks a spectroscopic
determination of [Fe/H] and only a photometric estimate is available.
While the model comparisons above suggest a metallicity fairly close
to that estimate, the solutions are not unique. We find that it is
possible to obtain similarly good fits for somewhat lower values of
[Fe/H], with the stars being at the end of the main-sequence phase
rather than past the ``blue hook''. We illustrate this with a third
set of models from the Victoria-Regina series
\citep{VandenBerg:06}. These calculations use a different description
of overshooting based on a parametrized version of the Roxburgh
criterion \citep{Roxburgh:78, Baker:87, Roxburgh:89}, in which the
effect is assumed to ramp up between masses of 1.15 and
1.70\,$M_{\sun}$ for compositions near solar.
Figure~\ref{fig:victoria} (top) shows a best-fit isochrone in the
$\log g$ vs.\ $T_{\rm eff}$ diagram that places the stars in the
Hertzprung gap, but as before it requires a mass ratio very near unity
($q = 1.0008$) at odds with the measured value at the 4.8$\sigma$
level. This best fit corresponds to an age of 2.9 Gyr and ${\rm
[Fe/H]} = -0.20$. In the bottom panel an equally good fit is achieved
for an age of 2.3 Gyr and ${\rm [Fe/H]} = -0.30$ that accommodates the
stars on the main sequence with the degree of overshooting prescribed
in the models. However, even here theory would require the masses to
be nearly the same ($q = 1.0169$) to satisfy the age constraint, which
is still in disagreement with the spectroscopic value at the
3.4$\sigma$ level.  Perhaps as importantly, the measured radii or
temperatures are not reproduced at the measured masses (see
Figure~\ref{fig:victoria}). The ages required to match them are again
younger for the more massive component, by about the same amount as in
the post-main-sequence solution.

A similar exercise with the PARSEC models from the Padova group
\citep{Bressan:13}, which employ yet another prescription for
overshooting, again allows two qualitatively different solutions. The
post-main-sequence scenario strongly disagrees with the measured mass
ratio, and a lower metallicity scenario \citep[$Z = 0.0085$, or ${\rm
[Fe/H]} = -0.25$ for $Z_{\sun} = 0.01524$;][]{Caffau:11} with the
stars at the end of the main-sequence still requires a mass ratio of
$q = 1.0153$ that is lower than our spectroscopic result at the
3.5$\sigma$ level.  Thus, all models seem to fail to match the
observations for \aq\ at a single age, pointing to a fundamental
problem with theory likely related to overshooting.

\section{Discussion}
\label{sec:discussion}

The most common way in which the degree of overshooting has been
calibrated in stellar evolution models is by means of star clusters,
and in particular through the comparison of isochrones to the
blue-hook region in the color-magnitude diagram (CMD). The drawbacks
are that this can be affected by contamination of the CMD by field
stars, unresolved binary systems, or variable stars, by uncertainties
in the chemical composition of the cluster, and even by systematics
from the color/temperature transformations.  Furthermore, the main
property of stars --- their absolute mass --- is generally not known
for any object along the CMD of the clusters most frequently used for
this type of comparison.  An alternate way of calibrating $\alpha_{\rm
ov}$ is by means of eclipsing binary systems, where masses and radii
are precisely known \citep[see, e.g.,][]{Ribas:00}. With this method
the latter authors found a relatively strong mass (and possibly
metallicity) dependence of overshooting for stars in the
2--12\,$M_{\sun}$ range, although subsequent studies reported the mass
dependence to be less pronounced and more uncertain \citep{Claret:07}.

The special location of \aq\ in the H-R diagram makes it a uniquely
sensitive test of convective core overshooting in current models of
stellar evolution. As shown above, the measured mass ratio is
different enough from unity that stars with radii and temperatures as
similar as they are in this system cannot possibly be in the
post-main-sequence phase. This constitutes clear evidence that mixing
beyond the core (overshooting) is required.  While early support for
the need of extra mixing was shown nearly 25 years ago by
\cite{Andersen:90} from the inordinately large number of B- and A-type
eclipsing systems that appeared to be in the Hertzprung gap, the
present system represents a particularly compelling demonstration.
However, \aq\ shows an additional problem, which is that even with the
inclusion of overshooting and the freedom to adjust the metallicity in
the models so as to accommodate the stars at the very end of the
main-sequence phase, current calculations are still unable to match
the well measured radius and temperature difference at the measured
masses.  Theory requires the stars to have masses much more similar to
each other ($q \approx 1.016$) than they are observed to be ($q =
1.054 \pm 0.011$), a difference that seems beyond reasonable
observational uncertainties.

This problem manifests itself also as an age discrepancy when
attempting to fit models to each component separately: the more
massive star appears systematically younger. Our comparisons in the
previous section indicate that this difficulty is common to all models
(and the age difference does not change much compared to the alternate
post-main-sequence fit): the age difference is about 0.5 Gyr for the
Granada models, 0.45 Gyr for the Yonsei-Yale models, 0.30 Gyr for the
Victoria-Regina models, and 0.40 Gyr for the Padova models, all
corresponding to 10--15\% of the mean evolutionary age of the binary.
Experiments with the Granada models in which we varied not only
$\alpha_{\rm ov}$ but also the mixing length parameter $\alpha_{\rm
ML}$ independently for each star, for different trial values of $Z$,
gave mean ages ranging from 3.1 to 3.5~Gyr, but did not improve the
situation regarding the age difference. We consistently found the
predicted age for the more massive star to be younger than the other
component.

\begin{figure}
\epsscale{1.1}
\plotone{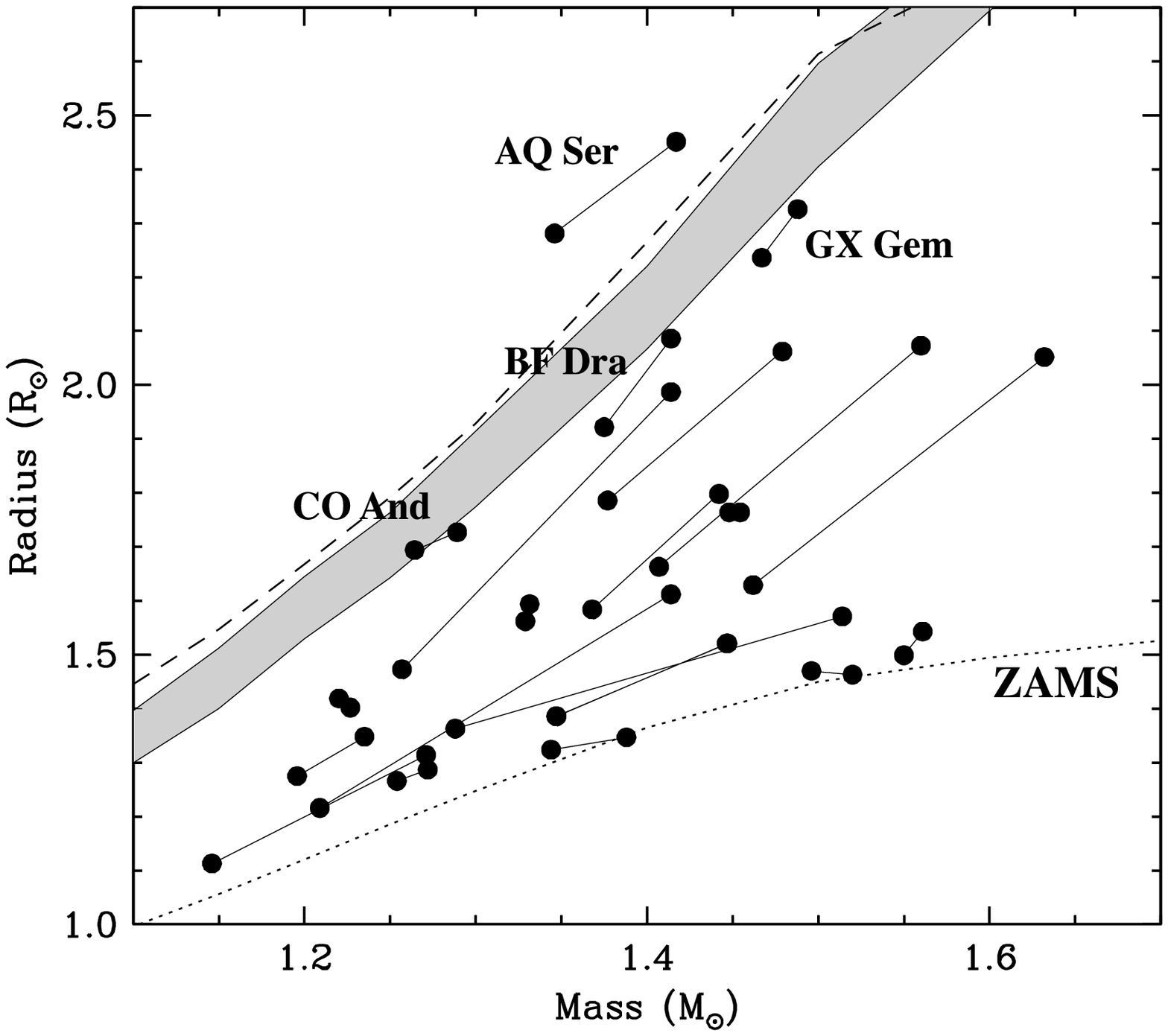}
\figcaption[]{
Masses and radii for all eclipsing binaries with accurately known
parameters (relative uncertainties in $M$ and $R$ less than 3\%) from
\cite{Torres:10a}, supplemented with measurements for CO\,And
\citep{Lacy:10}, BF\,Dra \citep{Lacy:12}, and \aq\ from the present
work. The primary and secondary stars in each system are connected
with a line. The shaded area corresponds to the blue hook region for
solar metallicity, according to the Yonsei-Yale models. The dashed
line represents the upper envelope of this region for a different
metallicity of ${\rm Fe/H]} = -0.20$ near that of \aq, and the dotted
line at the bottom corresponds to the zero-age main sequence (ZAMS).
\label{fig:mrdiagram}}
\end{figure}

Similar age discrepancies in the same direction as we see were pointed
out by \cite{Clausen:10} for several other well-measured F-type
eclipsing systems with unequal masses in the range from 1.15 to
1.70\,$M_{\sun}$. This is roughly the interval in which the models
ramp up the strength of the overshooting, and is also approximately
the mass range in which stars transition from having their energy
production dominated by the p-p chain to the CNO cycle.  The four
systems studied by \cite{Clausen:10} for which age differences were
noted are GX\,Gem, BW\,Aqr, V442\,Cyg, and BK\,Peg. Of these, the
first is the most similar to \aq\ in terms of its evolutionary
state. It is located just prior to the blue hook in the H-R diagram
according to the models, although the component masses are more
similar to each other than those in the \aq\ system, so the age
difference is less significant. Two other systems studied by us more
recently are also quite near the end of the main sequence: CO\,And
\citep{Lacy:10} and BF\,Dra \citep{Lacy:12}.  Although age anomalies
were not mentioned in the original investigations of these binaries, a
closer examination shows that both systems display age discrepancies
similar to those seen previously, with the more massive component
appearing younger. For CO\,And the age difference is about 0.3~Gyr
(8\%), and for BF\,Dra it is only $\sim$0.1~Gyr (4\%), but still in
the same direction. It is clear from these observations that a serious
deficiency has been uncovered in current stellar evolution models for
this mass range, which has not previously received much attention
beyond the work of \cite{Clausen:10}.

Figure~\ref{fig:mrdiagram} shows the location in the mass/radius
diagram of all well-measured eclipsing binary systems studied by
\cite{Clausen:10}, as well as others from \cite{Torres:10a} having
both components in the 1.15--1.70\,$M_{\sun}$ range. To these we added
CO\,And, BF\,Dra, and \aq. The shaded area represents the region of
the blue hook for solar composition, according to the Yonsei-Yale
models, and an increase in the overshooting parameter would shift this
region upward.  \aq\ is seen to be the most evolved system in this
mass range, which perhaps explains the larger age discrepancy noted
earlier.

Given that overshooting has a direct impact on evolution timescales,
particularly for main-sequence stars in the more advanced stages, it
is natural to suspect that the simplified treatment of this phenomenon
in current models has something to do with their difficulty in
matching the measured properties of binaries at a single age. However,
from our tests with \aq\ the explanation does not appear to be a
simple difference in $\alpha_{\rm ov}$ for the two components, and may
be more complex, involving, e.g., a dependence of overshooting on the
state of evolution, in addition to mass and metallicity. To our
knowledge the discrepancies highlighted by \aq\ and the other systems
mentioned above have not been investigated in detail for more massive
stars. Although such a study is beyond the scope of the present work,
it could well provide important clues about the nature of what we
consider one of the outstanding problems of stellar evolution theory.

\acknowledgements

We thank
P.\ Berlind,
M.\ Calkins,
G.\ Esquerdo,
D.\ W.\ Latham, and
R.\ P.\ Stefanik
for their assistance in obtaining the spectra of \aq, and R.\ J.\
Davis for maintaining the echelle database at the Harvard-Smithsonian
Center for Astrophysics. We also thank Dr.\ A.\ William (Bill) Neely,
who operates and maintains the NFO WebScope, and who handles
preliminary processing of the images and their distribution. We are
grateful to the anonymous referee for his/her helpful comments. GT
acknowledges partial support from this work from NSF grant
AST-1007992.  LPRV gratefully acknowledges partial support from the
Brazilian agencies CNPq, FAPEMIG, and CAPES.

\end{document}